\documentclass[pdflatex,sn-mathphys-num]{sn-jnl}
\usepackage[T1]{fontenc}
\usepackage{anyfontsize}
\usepackage{graphicx}
\usepackage{amsmath,amssymb,amsfonts}
\usepackage{amsthm}
\usepackage{mathtools}
\usepackage[title]{appendix}
\usepackage{xcolor}
\usepackage{textcomp}
\usepackage{booktabs}
\usepackage{subcaption}
\usepackage{setspace}
\usepackage[acronym,shortcuts]{glossaries}
\usepackage{hyperref}

\newcommand\ce[1]{\ensuremath{\mathrm{#1}}}

\newcommand\surfacemi[1]{$(#1)$}
\newcommand\directionmi[1]{$[#1]$}
\newcommand\surfacesmi[1]{$\{ #1 \}$}

\newcommand\pkgname{\texttt{SALAMI}}

\glsdisablehyper

\begin{document}

\title[Revealing low-energy surfaces of multinary compounds by controlling surface coordination environments]{Revealing low-energy surfaces of multinary compounds by controlling surface coordination environments}

\author*[1,2]{\fnm{Weihang} \sur{Xie}}\email{wxie@u.nus.edu}

\author[3,4]{\fnm{Harshan Reddy} \sur{Gopidi}}

\author[1]{\fnm{Zhengyu} \sur{Liu}}

\author[2]{\fnm{Romain} \sur{Claes}}

\author[2]{\fnm{Alexander G.} \sur{Squires}}

\author[5]{\fnm{Keith T.} \sur{Butler}}\email{k.t.butler@ucl.ac.uk}

\author[2]{\fnm{David O.} \sur{Scanlon}}\email{d.o.scanlon@bham.ac.uk}

\author[1,3,4]{\fnm{Pieremanuele} \sur{Canepa}}\email{pcanepa@uh.edu}

\affil*[1]{\orgdiv{Department of Materials Science and Engineering}, \orgname{National University of Singapore}, \orgaddress{\street{9 Engineering Drive 1}, \city{Singapore}, \postcode{117575}, \country{Singapore}}}
\affil[2]{\orgdiv{Department of Chemistry}, \orgname{University of Birmingham}, \orgaddress{\street{R35 University of Birmingham}, \city{Birmingham}, \postcode{B15 2TT}, \country{United Kingdom}}}
\affil[3]{\orgdiv{Department of Electrical and Computer Engineering}, \orgname{University of Houston}, \orgaddress{\street{4226 Martin Luther King Boulevard}, \city{Houston}, \postcode{TX 77204}, \state{Texas}, \country{United States}}}
\affil[4]{\orgname{Texas Center for Superconductivity}, \orgaddress{\street{3369 Cullen Blvd}, \city{Houston}, \postcode{TX 77204}, \state{Texas}, \country{United States}}}
\affil[5]{\orgdiv{Department of Chemistry}, \orgname{University College London}, \orgaddress{\street{Kathleen Lonsdale Building, Gower Place}, \city{London}, \postcode{WC1E 6BT}, \country{United Kingdom}}}

\abstract{
When modeling surfaces of multinary compounds, conventional cleavage planes often cut through strongly bonded polyhedra, resulting in unphysical surface energies.
Here, we introduce \pkgname{} (Symmetric Atomic Layers for Arbitrary Multinary Interfaces), a \texttt{Python} package that generates symmetric, charge-neutral, dipole-free, and low-energy slab models for multinary compounds.
\pkgname{} performs combinatorial searches to selectively remove surface atoms and generate corrugated terminations that preserve optimal coordination environments.
We applied this workflow to all symmetrically inequivalent crystallographic orientations with Miller indices up to 2 for two prototypical structures: the solid-state electrolyte \ce{Li_3PS_4} and the transparent conducting oxide \ce{ZnSb_2O_6}.
Density functional theory calculations reveal that \ce{Li_3PS_4} must preserve all \ce{PS_4} units to achieve the minimum surface energy.
For \ce{ZnSb_2O_6}, low-energy surfaces are achieved by partial undercoordination of surface \ce{Sb} atoms to \ce{SbO_5} or \ce{SbO_4} from the bulk \ce{SbO_6}, depending on the surface orientations.
Compared to surface models generated with unconstrained coordination, applying constraints to achieve optimal local coordination environments significantly lowers surface energies, shrinking the volume of the predicted Wulff shape by approximately 20\%.
Our results demonstrate that meticulous control of local coordination environments is necessary for accurately predicting the surface energetics of multinary compounds.
}


\maketitle

\section{Introduction}\label{sec:intro}
\noindent Accurate atomistic modeling of solid surfaces is a prerequisite for understanding interfacial phenomena in catalysis~\cite{allenElectronicStructuresSilver2011,brlecUnderstandingPhotocatalyticActivity2022},
thin-film engineering~\cite{wuEnhancedPhotoresponseFeS22016,hobsonIsotypeHeterojunctionSolar2020},
and semiconductor band engineering~\cite{scanlonBandAlignmentRutile2013,deacon-smithInterlayerCationExchange2014,donMultiPhaseSputteredTiO2Induced2023}.
The abrupt termination of a crystal at a surface causes strong deviations from the bulk properties, leading to unique properties, such as: electronic structure~\cite{brlecUnderstandingElectronicStructure2023},
reactivity~\cite{sibug-torresTransientAuClAdlayers2025},
thermodynamic instabilities, and others, which manifest differently from their bulk analogues~\cite{xieEffectsGrainBoundaries2024,logsdailBulkIonizationPotentials2014,canepaParticleMorphologyLithium2018}.
Studying these interfacial phenomena at the atomic scale requires appropriate models for surfaces.

Computationally, the surface is usually modeled by a periodic slab model, composed of several layers of bulk structure, and a sufficient thickness of vacuum to prevent interactions between periodic images~\cite{brlecSurfaxeSystematicSurface2021}.
The surface energy, $\gamma$, arising from broken bonds when cleaving the surface, can be calculated from the excess energy of the slab model compared to the bulk, as defined in \autoref{eq:gamma}~\cite{canepaParticleMorphologyLithium2018,butlerDesigningInterfacesEnergy2019},
\begin{equation}
 \label{eq:gamma}
 \gamma = \frac{1}{2S}\cdot \left[E_\mathrm{slab}-N_\mathrm{slab}E_\mathrm{bulk}-\displaystyle\sum_i ^\mathrm{species}\Delta n_i\mu_i \right]\, ,
\end{equation}
where $N_\mathrm{slab}$ is the number of formula units in the slab model,
$E_\mathrm{bulk}$ is the total energy of the bulk structure per formula unit,
$E_\mathrm{slab}$ is the total energy of the slab model.
$S$ is the area of the surface.
The factor of 2 corresponds to the two exposed surfaces in the slab model.
This requires that the top and bottom surfaces are identical~\cite{tranSurfaceEnergiesElemental2016}.
Otherwise, the resulting $\gamma$ becomes the average surface energy of two different surfaces, which is ambiguous.
For nonstoichiometric slabs, $\mu_i$ is the chemical potential of the species \textit{i}, which is derived from the phase diagram.
$\Delta n_i$ is the amount of off-stoichiometry for species \textit{i}.
All slab models considered in this work are stoichiometric, so that the chemical-potential term vanishes.

\begin{figure}[htbp!]
 \centering
 \includegraphics[width=0.7\textwidth]{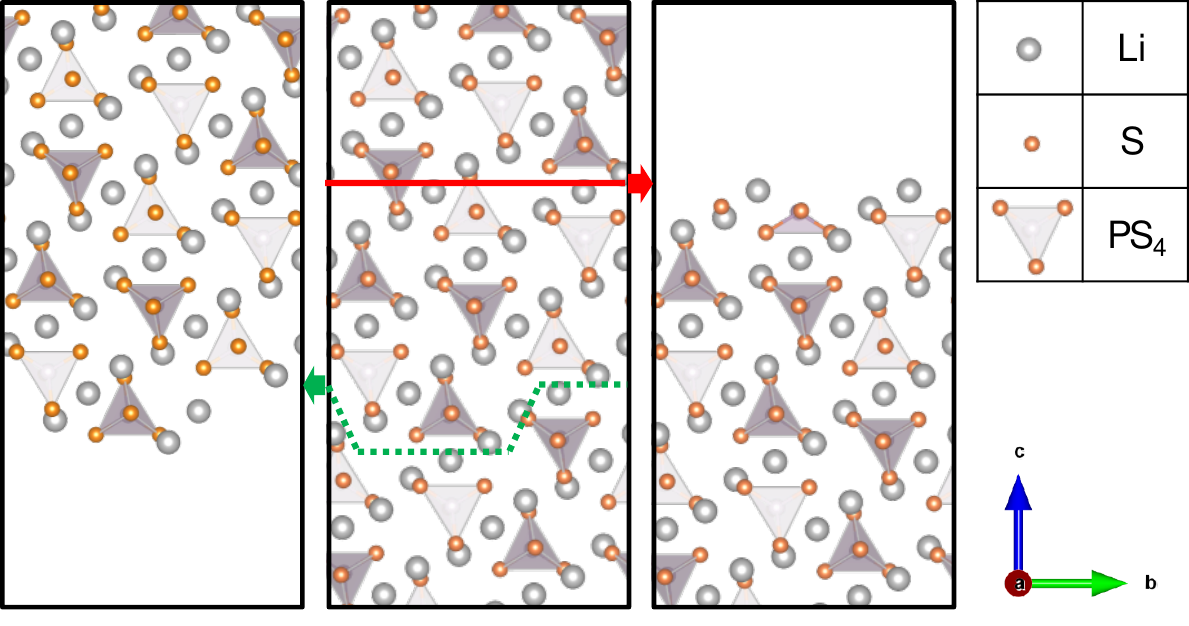}
 \caption[Cutting paths for $\beta$-\ce{Li_3PS_4}]{The \ce{Li_3PS_4} crystal in the \directionmi{011} direction is shown in the middle.
 The $c$-direction is set to \directionmi{011} direction.
 A conventional cleavage plane (red line) breaks the polyanion group, leading to high surface energy.
 The resulting surface morphology is shown in the structure on the right.
 On the \directionmi{011} direction, every cleavage plane (regardless of vertical displacement) breaks at least one \ce{PS_4} tetrahedron.
 Therefore, applying an optimal coordination environment, resulting in the corrugated surface on the left, is required to generate a low-energy \surfacemi{011} surface.
 }
 \label{fig:cuttingpath}
\end{figure}

While the conventional method of planar cleavage is generally sufficient to generate surface terminations for elemental or binary crystals, the inhomogeneity of bond strengths in multinary crystalline compounds complicates this approach significantly.
To demonstrate the challenges associated with constructing representative surface models for multinary systems, we investigate two prototype materials: the solid electrolyte $\beta$-\ce{Li_3PS_4} and the transparent conducting oxide \ce{ZnSb_2O_6}.
Specifically, \ce{Li_3PS_4} features isolated, strongly bonded polyanionic (\ce{PS_4^{3-}}) units. The \ce{P-S} bonds are stronger than the \ce{Li-S} bonds~\cite{culverEvidenceSolidElectrolyteInductive2020,wangBalancedSpHybridization2025,sunEnhancedIonicConductivity2022}, inferred from the higher diatomic bond dissociation energy of \ce{P-S} relative to \ce{Li-S}~\cite{haynesCRCHandbookChemistry2016}.
The inductive effect of the \ce{PS_4} group further weakens the \ce{Li-S} interactions~\cite{culverEvidenceSolidElectrolyteInductive2020}, evidenced by high ionic conductivity of Li~\cite{dietrichLithiumIonConductivity2017}.
High-energy bonds (such as the \ce{P-S} bond) can be cut by a cleavage plane (\textit{e.g.} red lines in \autoref{fig:cuttingpath}), leading to a slab model with high surface energy.
Minimizing the surface energy requires preserving optimal coordination environments that leave the \ce{PS_4} polyanions intact, thus generating a corrugated termination as indicated by the green line in \autoref{fig:cuttingpath}.
Different from isolated polyanions in \ce{Li_3PS_4}, \ce{ZnSb_2O_6} displays a continuous 3D network of corner- and edge-sharing \ce{SbO_6} octahedra.
Therefore, it is geometrically impossible to define a flat cleavage path that preserves all \ce{SbO_6} units while maintaining stoichiometry.
Constructing low-energy surface models for \ce{ZnSb_2O_6} requires partial undercoordination, which is detailed in the Results sections.

In this work, we studied ``representative'' slab models of \ce{Li_3PS_4} and \ce{ZnSb_2O_6} that satisfy three criteria:
\begin{enumerate}
\item Symmetric, \textit{i.e.}, the two exposed surfaces are identical. This also ensures the slab is a Tasker type 2~\cite{taskerStabilityIonicCrystal1979} slab with no surface dipoles perpendicular to the surface cut.
\item Stoichiometric, which establishes a baseline for surface energy calculations and avoids the ambiguity of chemical potentials. The stoichiometry of a slab also guarantees charge-neutrality.
\item Satisfy specific constraints on the coordination numbers of surface atoms. Enforcing these constraints achieves an optimal local coordination environment, which avoids high-energy surfaces (for example, keeping all \ce{PS_4} polyanions intact in \ce{Li_3PS_4}).
\end{enumerate}

To our knowledge, there are few automatic workflows available for identifying slab models that satisfy the criteria discussed above for multinary compounds.
While standard high-throughput libraries for surface generation, such as the Surfaxe package~\cite{brlecSurfaxeSystematicSurface2021} and the \texttt{pymatgen} module~\cite{sunEfficientCreationConvergence2013,tranSurfaceEnergiesElemental2016}, offer robust tools for slab generation---including options to symmetrize slabs by removing surface atoms or to fix broken polyanions by shifting atoms across the vacuum gap---these operations often either break the stoichiometry or the symmetry between the top and bottom terminations.
Recent developments, including the \texttt{polycleaver} package~\cite{mates-torresUnlockingSurfaceChemistry2024}, have addressed the heterogeneity of bond energy by preserving polyanionic frameworks during cleavage.
However, to our knowledge, there is no existing workflow that explicitly controls the surface coordination environment through systematic atom trimming to achieve fully symmetric and stoichiometric surfaces for multinary compounds.
As a result, current studies on surfaces of multinary compounds are either limited to very few special terminations where a cleavage plane coincidentally generates two identical surfaces and avoids breaking high-energy bonds, or rely on manual labor to reconstruct the surface, which is a tedious task~\cite{rohrbachInitioStudy00012004,canepaParticleMorphologyLithium2018}.
This limits the correct identification of the Wulff shape and leads to overestimation of surface energy.

To overcome this limitation, we present an automated workflow to preserve optimal coordination environments, generating symmetric, nonpolar, and low-surface-energy slab models for multinary compounds, implemented in the \texttt{Python} package \pkgname{} (Symmetric Atomic Layers for Arbitrary Multinary Interfaces)\footnote{The code is open source at \url{https://github.com/caneparesearch/salami}}.
We demonstrate the complete workflow of \pkgname{} by identifying low-energy surfaces fulfilling the important criteria exposed above of \ce{Li_3PS_4} and \ce{ZnSb_2O_6}.
Through these two examples, we highlight different principles for surface stabilization: strongly bonded units like \ce{PS_4} must be preserved in \ce{Li_3PS_4}, whereas interconnected units like \ce{SbO_6} often require partial undercoordination at the surface to satisfy local charge neutrality in \ce{ZnSb_2O_6}.

\section{Results}\label{sec2}

\subsection{Generating symmetric and stoichiometric slab models}\label{sec:workflow}

\begin{figure}[htbp]
 \centering
 \includegraphics[width=0.8\textwidth]{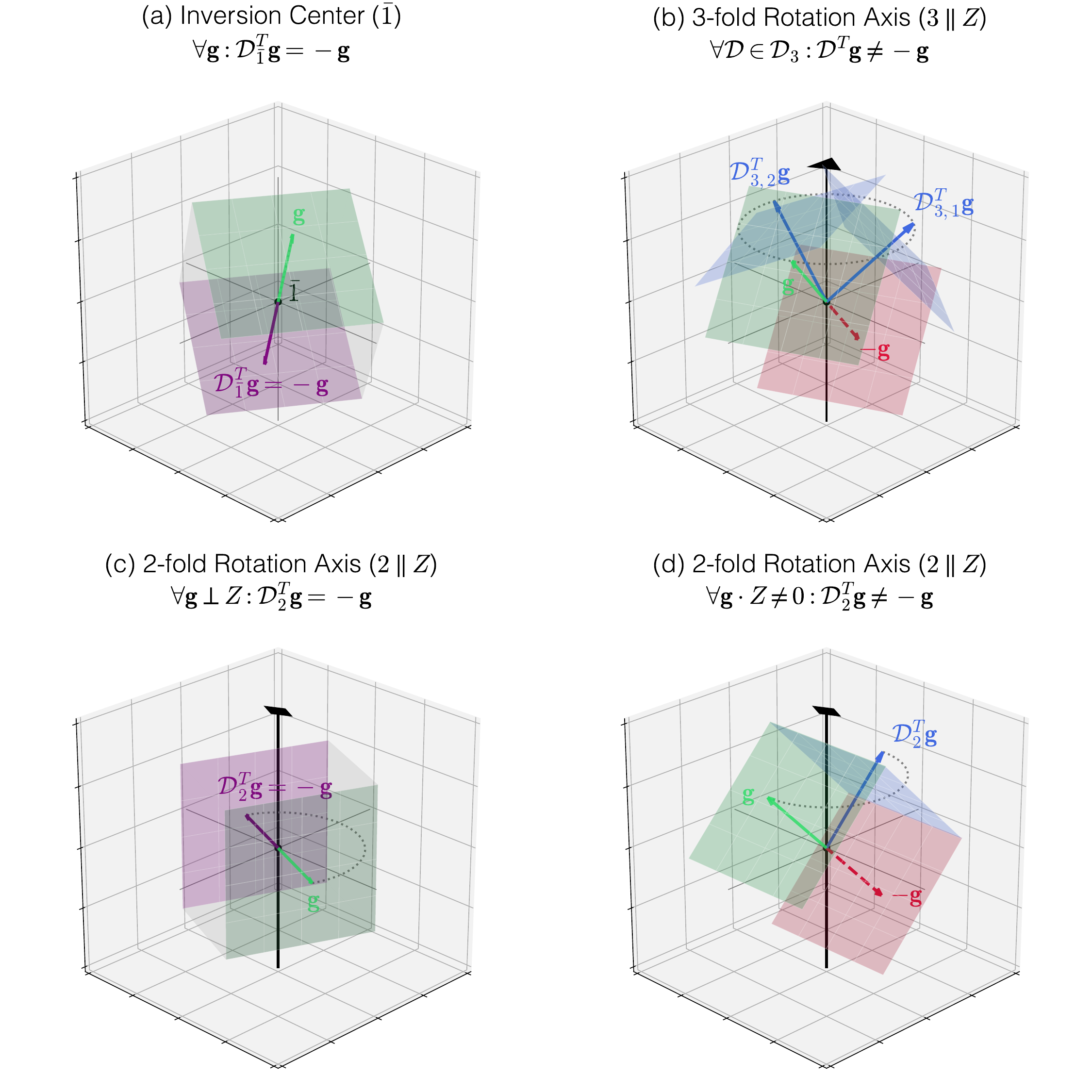}
 \caption{Geometric conditions for generating symmetric slabs via point-group operations.
 The top surface normal $\mathbf{g}$ (green), target bottom surface normal $-\mathbf{g}$ (red), and symmetry-mapped normal $\mathcal{D}^{T}\mathbf{g}$ (blue) are projected in real space.
 Purple vectors denote successful mapping where $\mathcal{D}^{T}\mathbf{g} = -\mathbf{g}$ is satisfied.
 (a) An inversion center ($\bar{1}$) maps any $\mathbf{g}$ to $-\mathbf{g}$.
 (b) A 3-fold rotation axis on $z$-direction ($3 \parallel Z$) fails to map any $\mathbf{g}$ to $-\mathbf{g}$, as $\mathcal{D}^{T}\mathbf{g} \neq -\mathbf{g}$ for any $\mathbf{g}$.
 (c) A 2-fold rotation axis on $z$-direction ($2 \parallel Z$): when $\mathbf{g}$ is perpendicular to the rotation axis ($\mathbf{g} \perp Z$), $\mathcal{D}^{T}_2\mathbf{g} = -\mathbf{g}$ is satisfied.
 (d) The 2-fold operation fails to map $\mathbf{g}$ to $-\mathbf{g}$ if $\mathbf{g}$ is not perpendicular to the rotation axis.
 }
 \label{fig:symmetry_slabs}
\end{figure}

\noindent Generating a symmetric slab model along an arbitrary $(hkl)$ plane is not always possible if the crystal is not centrosymmetric.
Cleaving a bulk crystal along an $(hkl)$ plane exposes a top surface with an outward normal vector parallel to the reciprocal lattice vector $\mathbf{g} = (h, k, l)^T$, and a bottom surface with the normal $\mathbf{g}' = -\mathbf{g} = (-h, -k, -l)^T$.
For the resulting slab to be symmetric, the bulk space group must contain at least one point-group operation---represented by a $3 \times 3$ matrix $\mathcal{D}$~\cite{degraefStructureMaterialsIntroduction2012}---that maps the top surface directly onto the bottom surface, \textit{i.e.}, this symmetry operation inverts the surface normal.
Here, \autoref{fig:symmetry_slabs} shows 4 examples to illustrate conditions for generating symmetric slabs via point-group operations.

In reciprocal space, $\mathbf{g}$ transforms under the real-space operation $\mathcal{D}$ according to $\mathbf{g}' = \mathcal{D}^{-T}\mathbf{g}$~\cite{mullerSymmetryRelationshipsCrystal2013}.
Equating the two expressions for $\mathbf{g}'$ yields $\mathcal{D}^{-T}\mathbf{g} = -\mathbf{g}$.
Multiplying both sides by the transpose matrix $\mathcal{D}^T$ removes the inverse operation and leads to an eigenvalue equation:
\begin{equation}
 \label{eq:eigen}
 \mathcal{D}^T\mathbf{g} = -\mathbf{g}.
\end{equation}

\autoref{eq:eigen} indicates that a symmetric slab can be constructed along the $(hkl)$ direction only if $\mathbf{g}$ is an eigenvector of $\mathcal{D}^T$ with an eigenvalue of $\lambda = -1$, \textit{i.e.}, the crystal possesses a symmetry operation that reverses the surface normal vector.
The eigenspace falls into four scenarios:
\begin{enumerate}
 \item \textbf{Center of inversion ($\bar{1}$):} When $\mathcal{D}$ represents an inversion operation ($\mathcal{D} = -I$), the eigenspace for $\lambda = -1$ spans all real-space directions, as an inversion flips all directions. Symmetric slabs can be generated along all $(hkl)$ directions.
 \item \textbf{Two-fold rotation ($2$):} When $\mathcal{D}$ defines a two-fold rotation, the eigenspace for $\lambda = -1$ is a 2D plane perpendicular to the rotation axis defined by $\mathcal{D}$ (a $180^\circ$ rotation reverses directions within the plane perpendicular to the rotation axis). Therefore, a symmetric slab can be generated if the surface normal $\mathbf{g}$ lies within this plane.
 \item \textbf{Mirror plane ($m$):} When $\mathcal{D}$ represents a mirror plane, the eigenspace for $\lambda = -1$ is a 1D line parallel to the normal of that mirror plane (a reflection reverses only the direction orthogonal to the mirror plane). Therefore, a symmetric slab can be generated if the surface normal $\mathbf{g}$ is perpendicular to a mirror plane of the crystal.
 \item \textbf{Four-fold improper rotation ($\bar{4}$):} When $\mathcal{D}$ defines a four-fold improper rotation, the eigenspace for $\lambda = -1$ is a 1D line parallel to the rotation axis, and a symmetric slab can be generated if the surface normal $\mathbf{g}$ is parallel to the $\bar{4}$ axis. A symmetric slab can also be generated if the surface normal $\mathbf{g}$ is  perpendicular to the $\bar{4}$ rotation axis, due to the inherent 2-fold rotation ($\bar{4}^2 = 2$, scenario 2).

\end{enumerate}

From the above analysis, symmetric slabs can be generated for any $(hkl)$ direction in centrosymmetric crystals, which contain an inversion center.
In non-centrosymmetric crystals, symmetric slabs are forbidden unless the target $(hkl)$ plane is parallel to a $2$-fold rotation axis, parallel to a mirror plane, or perpendicular to a $\bar{4}$ axis.
For example, in cubic systems with space group $P432$ (No.~207), symmetric slabs cannot be generated for arbitrary Miller indices, such as $(321)$.
In crystals belonging to space groups $P1$ (No.~1) or $P3$ (No.~143), no symmetry operations yield an eigenvalue of $\lambda = -1$ for any $\mathbf{g}$, making it difficult to generate symmetric slabs on any orientation unless the slab is significantly reconstructed.
Here, both $\beta$-\ce{Li_3PS_4} ($Pnma$, No.~62~\cite{hommaCrystalStructurePhase2011}) and \ce{ZnSb_2O_6} ($P4_2/mnm$, No.~136~\cite{jacksonComputationalPredictionExperimental2022}) are centrosymmetric crystals that satisfy \autoref{eq:eigen}.
For these \ce{Li_3PS_4} and \ce{ZnSb_2O_6}, we explore all symmetrically inequivalent orientations of surface cuts.

As illustrated in \autoref{fig:cuttingpath}, minimizing the surface energy of multinary compounds frequently requires constrained coordination to bypass strong bonds and preserve an optimal local coordination environment at the surface.
To automate the construction of these optimally coordinated surfaces, \pkgname{} trims atoms rigorously from an initial slab formed by a simple cleavage plane.
This trimming process is controlled by user-defined constraints on coordination numbers, which can be derived from empirical chemical knowledge or quantitative bonding analysis.
These constraints ensure that strong bonds are preserved, while weakly bonded atoms are removed to restore stoichiometry or charge neutrality.

\begin{figure}[htbp!]
 \centering
 \includegraphics[width=1\textwidth]{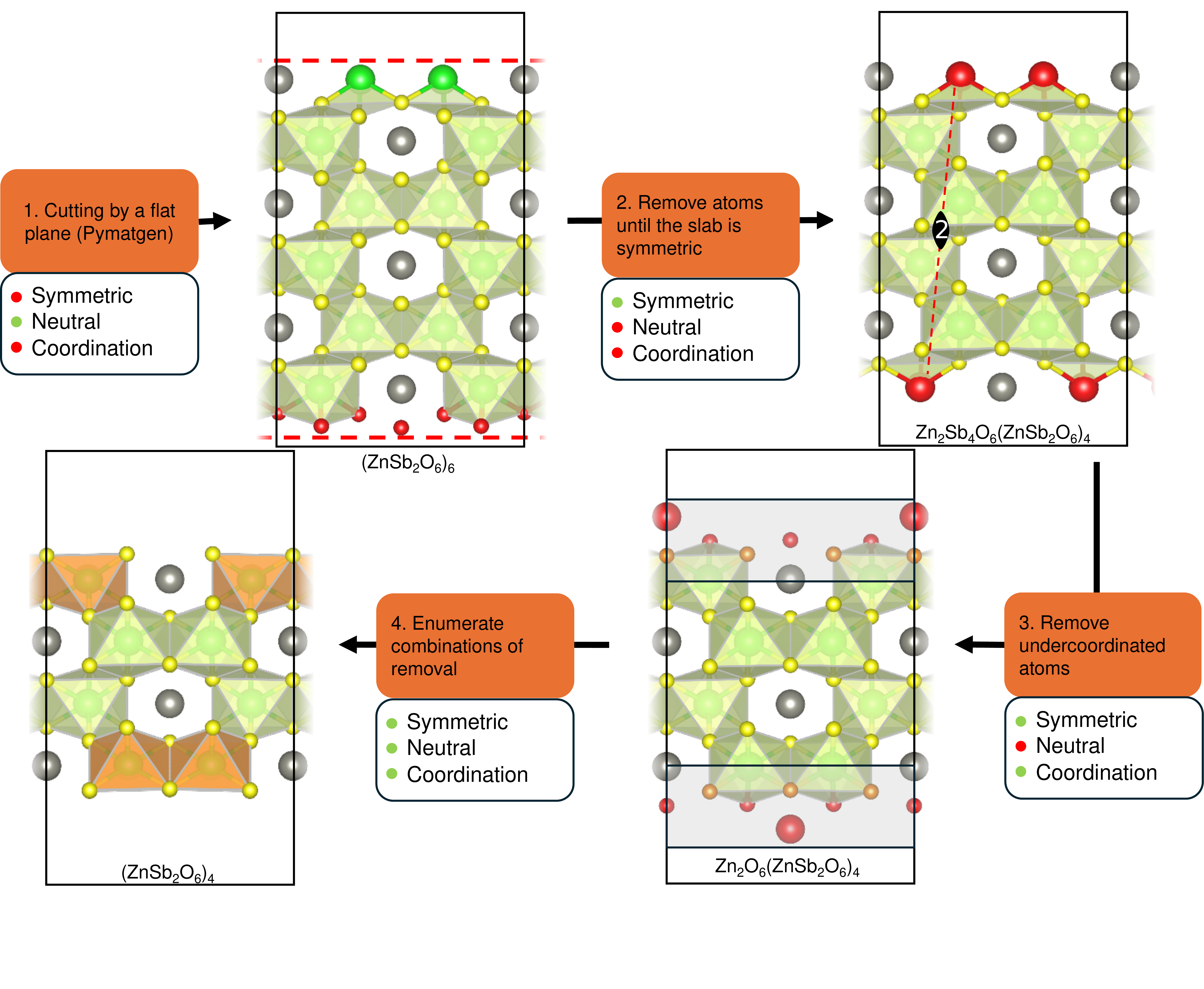}
 \caption[Workflow of \pkgname{} class]{Workflow of \pkgname{} on an exemplary \ce{ZnSb_2O_6} structure:
 Grey spheres are Zn atoms; green spheres are Sb atoms; yellow spheres are O atoms; lime octahedra are \ce{SbO_6} units; orange polyhedra are surface \ce{SbO_5} units. The black inset boxes display the criteria evaluation (green dot is satisfied, red dot is unsatisfied).
 \textbf{1}: Initial Cleavage: The bulk structure is cleaved via a cleavage plane (red dashed line), yielding a stoichiometric but asymmetric slab, \ce{(ZnSb_2O_6)_6}.
 \textbf{2}: Symmetrization: Atoms (marked in red in step 1) are removed from the bottom until the slab is symmetric, resulting in a non-stoichiometric intermediate \ce{Zn_2Sb_4O_6(ZnSb_2O_6)_4}. A 2-fold rotation axis is shown for better illustration.
 \textbf{3}: Coordination Trimming: Atoms violating minimum coordination constraints (marked in red in step 2) are symmetrically removed.
 \textbf{4}: Stoichiometric Recovery: Within a defined surface depth (grey shaded region), combinatorial removal of specific atoms (red spheres) is performed to restore exact stoichiometry and charge neutrality, leaving only optimally coordinated surface atoms (orange spheres), thereby yielding a valid slab with minimized surface energy.}
 \label{code:fig:slabworkflow}
\end{figure}

We demonstrate this constraint-driven workflow using \ce{ZnSb_2O_6} (\autoref{code:fig:slabworkflow}).
Here, the minimum allowed coordination number (min CN) is set to 5 for \ce{Sb} as an initial heuristic.
Additional constraints are applied to \ce{Zn} and \ce{O} atoms to prevent the formation of dangling bonds in the vacuum region, as detailed in \autoref{tab:coordinationworkflow}.
The physical rationale for establishing the actual optimal coordination limits for both \ce{ZnSb_2O_6} and \ce{Li_3PS_4} is further discussed in the next section.

Optimal coordination environments are achieved by removing surface atoms in \pkgname{}.
The removal process consists of four sequential stages.
First, an initial stoichiometric slab is generated via a cleavage plane.
This initial slab is not necessarily symmetric.
Subsequently, \pkgname{} removes atoms from the bottom surface until the symmetry of the slab is achieved.
This step usually breaks the stoichiometry and the charge-neutrality of the slab.
If the symmetry cannot be achieved on the given orientation, the slab is discarded.
In the third step, the algorithm targets and removes undercoordinated surface atoms that violate the predefined coordination constraints (\textit{e.g.}, an \ce{Sb} atom coordinated by fewer than 5 \ce{O} atoms).
Because the overall slab is already symmetric, this trimming is applied symmetrically to both the top and bottom surfaces to preserve the symmetry of the slab model.

Finally, \pkgname{} performs a combinatorial search to restore stoichiometry and charge neutrality.
The algorithm identifies ``surface atoms'' within a user-defined depth and enumerates all possible combinations of symmetrically removing $n$ atoms (starting from $n=1$) from the $m$ available surface atoms on both sides.
If no valid slab is found, $n$ is incrementally increased.
For instance, in Step 4 of \autoref{code:fig:slabworkflow}, achieving the stoichiometry requires the symmetric removal of 1 \ce{Zn} and 3 \ce{O} atoms from each side.
Given the available surface atoms (1 \ce{Zn} and 6 \ce{O}) within the search region, \pkgname{} evaluates $C_1^1 \times C_6^3 = 20$ possible removal combinations.
Each resulting configuration is evaluated against the predefined criteria (further discussed in \autoref{sec:package}).
Configurations that violate the coordination constraints (\textit{e.g.}, removing the orange, instead of red \ce{O} atoms in \autoref{code:fig:slabworkflow}, which breaks the coordination constraint of \ce{Sb}) are discarded.
Finally, representative slab models are filtered by their energies, detailed in \autoref{sec:package}.

\subsection{Iterative determination of optimal coordination environments}

\noindent While \pkgname{} automates the atom trimming of complex surfaces, surface energies of the resulting slab models heavily depend on the coordination constraints.
The optimal constraints for a new multinary compound are rarely known \textit{a priori}.
Identifying optimal coordination environments for a new multinary compound requires a systematic approach.
The process consists of three steps.
First, crystal orbital Hamiltonian population (COHP)~\cite{nelsonLOBSTERLocalOrbital2020,maintzLOBSTERToolExtract2016,deringerCrystalOrbitalHamilton2011,dronskowskiCrystalOrbitalHamilton1993} analysis identifies the strongest bonds that form the primary coordination polyhedra.
Second, Pauling's electrostatic valence principle determines the upper limit for preserving these polyhedra at the surface.
Finally, we iteratively evaluate coordination constraints below this upper limit to identify the optimal lower bound that yields the lowest surface energies.

\begin{figure}[htbp!]
 \centering
 \includegraphics[width=0.7\textwidth]{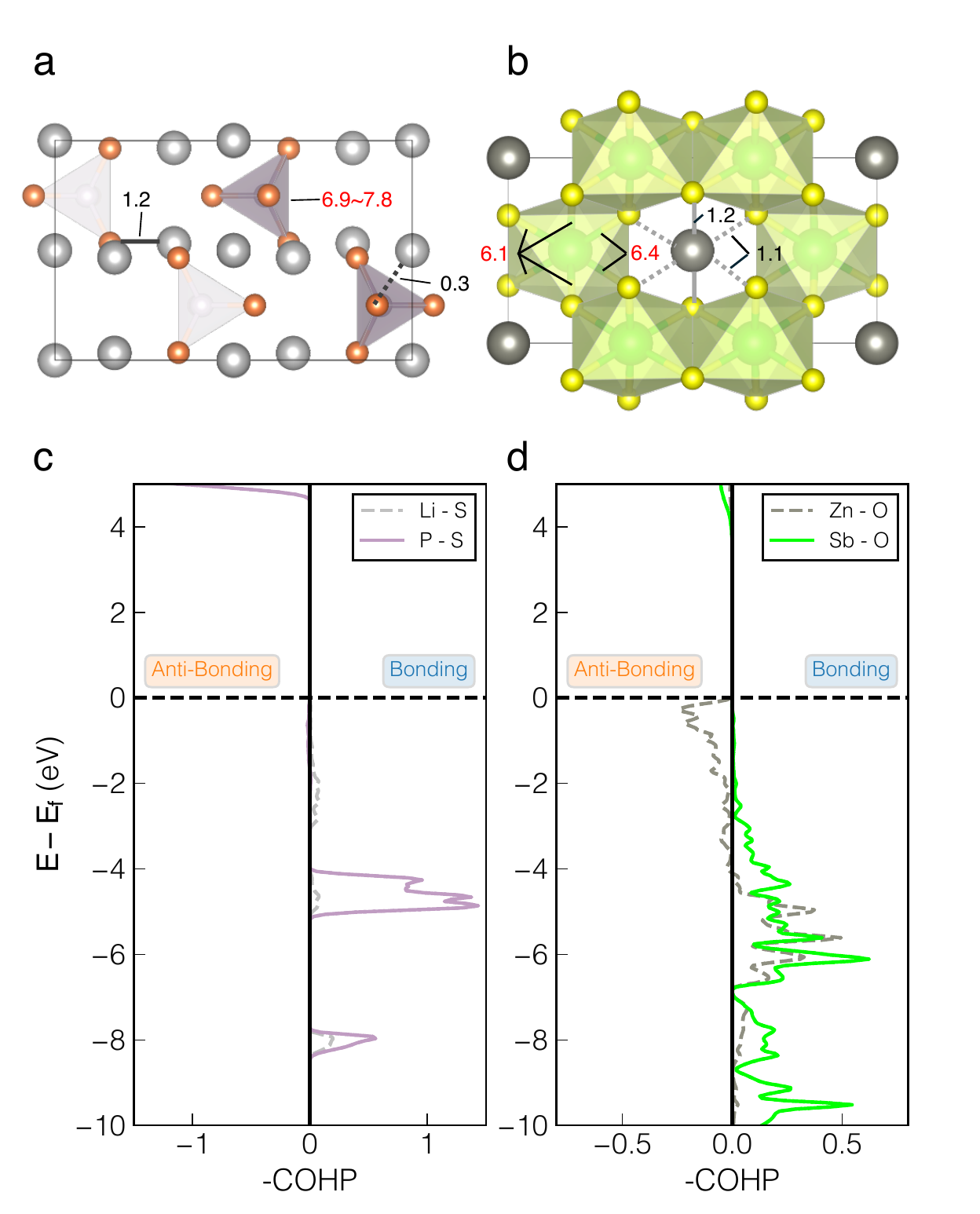}
 \caption{Bulk structures of \ce{Li_3PS_4} and \ce{ZnSb_2O_6} are shown in \textbf{a} and \textbf{b}, respectively. The corresponding COHP analyses are shown in \textbf{c} and \textbf{d}.
 \textbf{a}: The bulk structure of \ce{Li_3PS_4}. Grey tetrahedra are \ce{PS_4} units; silver spheres are Li atoms; orange spheres are S atoms. ICOHP values (in eV) are annotated on representative bonds, showing larger values for \ce{P-S} bonds (6.9--7.8) compared to \ce{Li-S} bonds (0.3--1.2). The primary coordination is identified as \ce{PS_4}.
 \textbf{b}: The bulk structure of \ce{ZnSb_2O_6}. Green octahedra are \ce{SbO_6} units; grey spheres are Zn atoms; yellow spheres are O atoms. Annotated ICOHP values (in eV) demonstrate higher bond strength for \ce{Sb-O} (6.1--6.4) than for \ce{Zn-O} (1.1--1.2). The primary coordination is identified as \ce{SbO_6}.
 \textbf{c}: The $\mathrm{-COHP}$ between \ce{P-S} is significantly higher than that of \ce{Li-S} in the valence band region ($E-E_f<0$), indicating that \ce{P-S} bonds are stronger than \ce{Li-S} bonds.
 \textbf{d}: The $\mathrm{-COHP}$ between \ce{Sb-O} is higher than that of \ce{Zn-O} in the valence band region. Additionally, the anti-bonding character (negative $\mathrm{-COHP}$) of \ce{Zn-O} near the valence band maximum (VBM) further decreases its bond strength.}
 \label{code:fig:lobster}
\end{figure}

To identify the primary coordination environments for \ce{Li_3PS_4} and \ce{ZnSb_2O_6}, we first investigated their bonding heterogeneities using the COHP analysis (\autoref{code:fig:lobster}).
A more negative integrated COHP value ($-\mathrm{ICOHP}$) of a bond indicates stronger bonding and a larger energy penalty upon bond cleavage.
In \ce{Li_3PS_4} (\autoref{code:fig:lobster}a,c), the $-\mathrm{ICOHP}$ values of \ce{P-S} bonds (6.9 to 7.8~eV) are larger than those of \ce{Li-S} bonds (0.3 to 1.2~eV).
In \ce{ZnSb_2O_6} (\autoref{code:fig:lobster}b,d), the $-\mathrm{ICOHP}$ of \ce{Sb-O} (6.1 to 6.4~eV) exceeds that of \ce{Zn-O} (1.1 to 1.2~eV) due to the antibonding interaction of \ce{Zn-O} near the valence band maximum.
Consequently, \ce{PS_4} and \ce{SbO_6} are identified as the primary coordination units for \ce{Li_3PS_4} and \ce{ZnSb_2O_6}, respectively.

Although it appears to be energetically favorable to preserve primary coordination units, the maximum coordination number achievable at the surface is limited by local charge neutrality.
We quantify this electrostatic imbalance using Pauling's electrostatic valence principle to establish the upper bound for surface coordination.
Consider a multinary compound with a general formula of $\ce{A_x B_y C_z}$, where \ce{A} and \ce{B} are cations with formal charges $a$ and $b$, and \ce{C} is an anion with formal charge $-c$ ($a,b,c>0$).
The possibility of maintaining the primary coordination environment $\ce{BC_{\beta }}$ ($\beta$ is the coordination number of \ce{B}) at the surface is governed by local charge neutrality.
Assuming a stoichiometric termination where each \ce{BC_{\beta }} polyhedron contributes $N_{\mathrm{term}}$ undercoordinated anion (\ce{C}) to the surface (generally, $N_{\mathrm{term}} = 1$), we first define the electrostatic valence deficiency $\Delta$ for undercoordinated anion \ce{C}:
\begin{equation}
 \Delta=|c|-\frac{b}{\beta}.
\end{equation}
This deficiency must be entirely compensated by the available $\ce{A}$ cations.
To evaluate this, we propose the surface coordination stability index ($\eta$), which serves as a simple heuristic representing the ratio of the electrostatic compensating charge provided by the available surface cations to the electrostatic valence deficiency created by the uncoordinated surface anions.
This index allows for a rapid evaluation on whether the stoichiometric amount of \ce{A} is sufficient:
\begin{equation}
 \eta = \frac{{x a}}{y N_{\mathrm{term}}\Delta}.
\end{equation}
Here, $\eta > 1$ serves as a heuristic condition for generating stoichiometric surfaces.

For \ce{Li_3PS_4}, $\eta=\frac{4}{N_{\mathrm{term}}}$, even in the extreme scenario where three coplanar S atoms in a \ce{PS_4} unit point outward ($N_{\mathrm{term}}=3$), it is possible to generate surfaces with all \ce{PS_4} coordination intact.
Conversely, for \ce{ZnSb_2O_6}, $\eta = 0.86 < 1$, indicating that full \ce{SbO_6} coordination cannot be maintained at the surface.
Detailed derivation is in \autoref{sec:si:pauling}.
Briefly, in an ideal stoichiometric \ce{ZnSb_2O_6} surface with intact \ce{SbO_6} octahedra, the ``outer'' oxygen atoms extending into the vacuum exhibit a valence deficiency that must be compensated by the available surface \ce{Zn^{2+}} cations.
However, the available \ce{Zn} cations are electrostatically insufficient to balance the cumulative deficiency of these terminal oxygens.
Consequently, the coordination number of \ce{Sb} has to be reduced to 5 (\ce{SbO_5}) or even 4 (\ce{SbO_4}), depending on the orientation of the surface.

\begin{figure}[htbp!]
 \centering
 \includegraphics[width=0.6\textwidth]{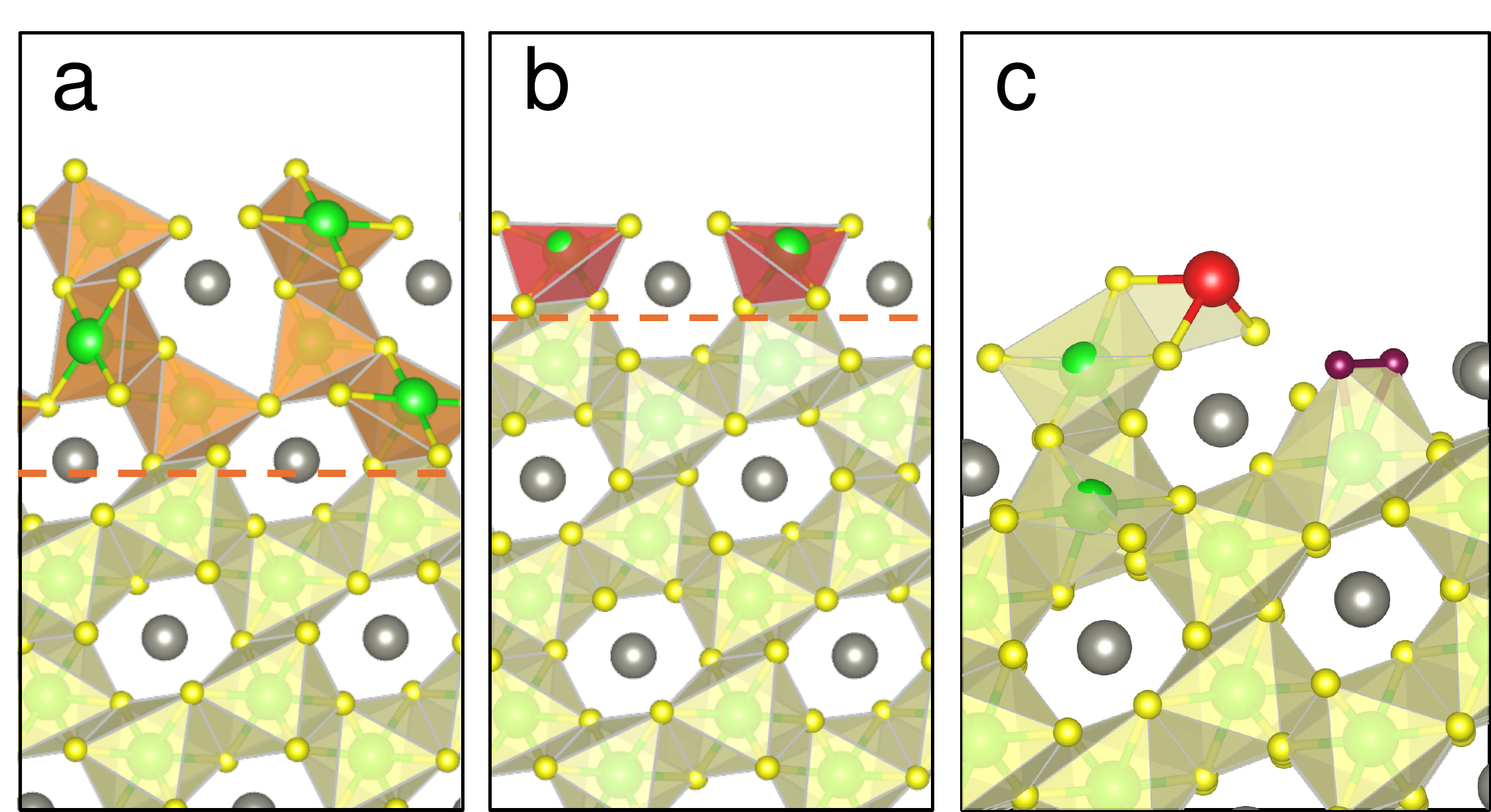}
 \caption{Over-coordination at \ce{ZnSb_2O_6} \surfacemi{101} surface before (\textbf{a}) and after (\textbf{b}) relaxation, and undercoordination at \ce{ZnSb_2O_6} \surfacemi{212} surface (\textbf{c}).
 \textbf{a}: Before relaxation, forcing a strict \ce{SbO_5} minimum coordination drives the algorithm to extract oxygen atoms that are deep inside the bulk, resulting in three layers of broken \ce{SbO_5} polyhedra (the polyhedra marked in orange above the red dashed line).
 \textbf{b}: After DFT relaxation, \ce{SbO_6} coordination is spontaneously restored in the subsurface bulk region via surface rumpling. \ce{O} atoms migrate inward to heal the bulk, leaving undercoordination to the outermost surface layer (\ce{SbO_4} marked in red above the red dashed line).
 \textbf{c}: A relaxed stoichiometric \surfacemi{212} surface model with unconstrained coordination (\ce{SbO_3}). Structural relaxation causes reduction of a \ce{Sb} atom (red sphere) and formation of an \ce{O-O} dimer (purple dumbbell).}
 \label{fig:12_101_016}
\end{figure}

Determining the optimal lower limit of the coordination number requires evaluating the energetic trade-off between preserving surface \ce{Sb-O} bonds and minimizing subsurface structural deformations.
Intuitively, enforcing a higher min CN, such as \ce{SbO_5} in \ce{ZnSb_2O_6} in \autoref{code:fig:slabworkflow}, appears to break fewer \ce{Sb-O} bonds at the surface.
However, satisfying these strict coordination constraints while restoring stoichiometry forces the algorithm to expand the search region deeper into the slab.
For example, on the \surfacemi{101} surface of \ce{ZnSb_2O_6}, \ce{O} atoms deep inside the bulk must be removed to maintain a minimum coordination of 5 (\ce{SbO_5}), as shown in \autoref{fig:12_101_016}a.

This deep extraction results in an initial surface exposing three consecutive layers of broken polyhedra above the bulk framework (\autoref{fig:12_101_016}a).
Subsequent DFT relaxation (\autoref{fig:12_101_016}b) reveals a spontaneous reconstruction driven by surface rumpling: subsurface \ce{O} atoms migrate inward to restore the bulk \ce{SbO_6} octahedra, leaving the outermost surface layer as \ce{SbO_4}.
This reconstruction indicates that the energy penalty for creating bulk vacancies exceeds that of surface undercoordination.
Therefore, rather than increasing the search depth to preserve \ce{SbO_5}, a more reasonable strategy is to relax the coordination constraints (\textit{e.g.}, permitting \ce{SbO_4}).

Although certain undercoordination can stabilize the surface, allowing arbitrary undercoordination leads to unphysical reconstructions.
For instance, using a cleavage plane and removing minimal surface atoms to achieve symmetry without applying coordination constraints, we generated a stoichiometric \surfacemi{212} surface model with \ce{SbO_3} units (\autoref{fig:12_101_016}c).
After DFT relaxation, Bader charge analysis indicates a localized reduction of the tri-coordinated \ce{Sb} sites from \ce{Sb^{5+}} to \ce{Sb^{3+}} (\autoref{sec:si:bader}).
As the stoichiometry and the charge neutrality are maintained, this reduction is offset by an oxidation in the neighboring oxygen atoms, leading to the formation of an \ce{O-O} dimer with a bond length of 1.6~\AA{}.
Therefore, the optimal coordination constraints for \ce{ZnSb_2O_6} are set to a min CN of either 4 (\ce{SbO_4}) or 5 (\ce{SbO_5}) for Sb, depending on the surface orientation.

For \ce{Li_3PS_4}, any broken \ce{PS_4} unit increases the surface energy.
Unlike the adaptable \ce{SbO_6} framework in \ce{ZnSb_2O_6}, the \ce{PS_4} polyanion is rigid compared to \ce{Li-S} bonds in \ce{Li_3PS_4}.
Chemically, the \ce{PS_4} tetrahedron is a polyanion with strong covalent \ce{P-S} bonds.
Cleaving these bonds destroys the $sp^3$ hybridization of phosphorus, creating high-energy dangling bonds.
Additionally, preserving \ce{PS_4} coordinations at the surface does not induce unphysical redox reactions.
The mobility of \ce{Li} and a higher cation density (three \ce{Li} per \ce{PS_4} compared to one \ce{Zn} per two \ce{SbO_6}) enable \ce{Li_3PS_4} to stabilize intact polyanions and mitigate local electrostatic imbalances.
Therefore, the optimal coordination constraint for \ce{Li_3PS_4} is to strictly preserve all \ce{PS_4} coordinations.

\subsection{Optimal coordination and surface energies}\label{sec:undercoordination}

\noindent To evaluate the energetic penalties associated with the undercoordination of strongly bonded polyhedra across various surface orientations, we quantify the effects of constrained coordination.
Using \pkgname{}, we compare slab models with unconstrained coordination---where cleavage arbitrarily truncates \ce{PS_4} tetrahedra in \ce{Li_3PS_4} or neglects \ce{Sb} coordination in \ce{ZnSb_2O_6}---against slabs with constrained coordination that avoid breaking high-energy bonds.
It is worth noting that a raw planar cleavage along certain orientations (such as \surfacemi{100}, \surfacemi{011}, \surfacemi{102}, and \surfacemi{111} in \ce{Li_3PS_4}) inevitably breaks \ce{PS_4} units.
However, the unconstrained coordination baseline in \pkgname{} yields fully intact polyhedra on these orientations due to the symmetrization process (Step 2 in \autoref{code:fig:slabworkflow}), which coincidentally drops undercoordinated atoms (detailed in \autoref{sec:si:flat}).
The specific constraints of coordination numbers for both scenarios are detailed in \autoref{tab:coordination}.
All generated slabs are symmetric, stoichiometric, and dipole-free.

\begin{figure}[htbp!]
\centering
\includegraphics[width=1\textwidth]{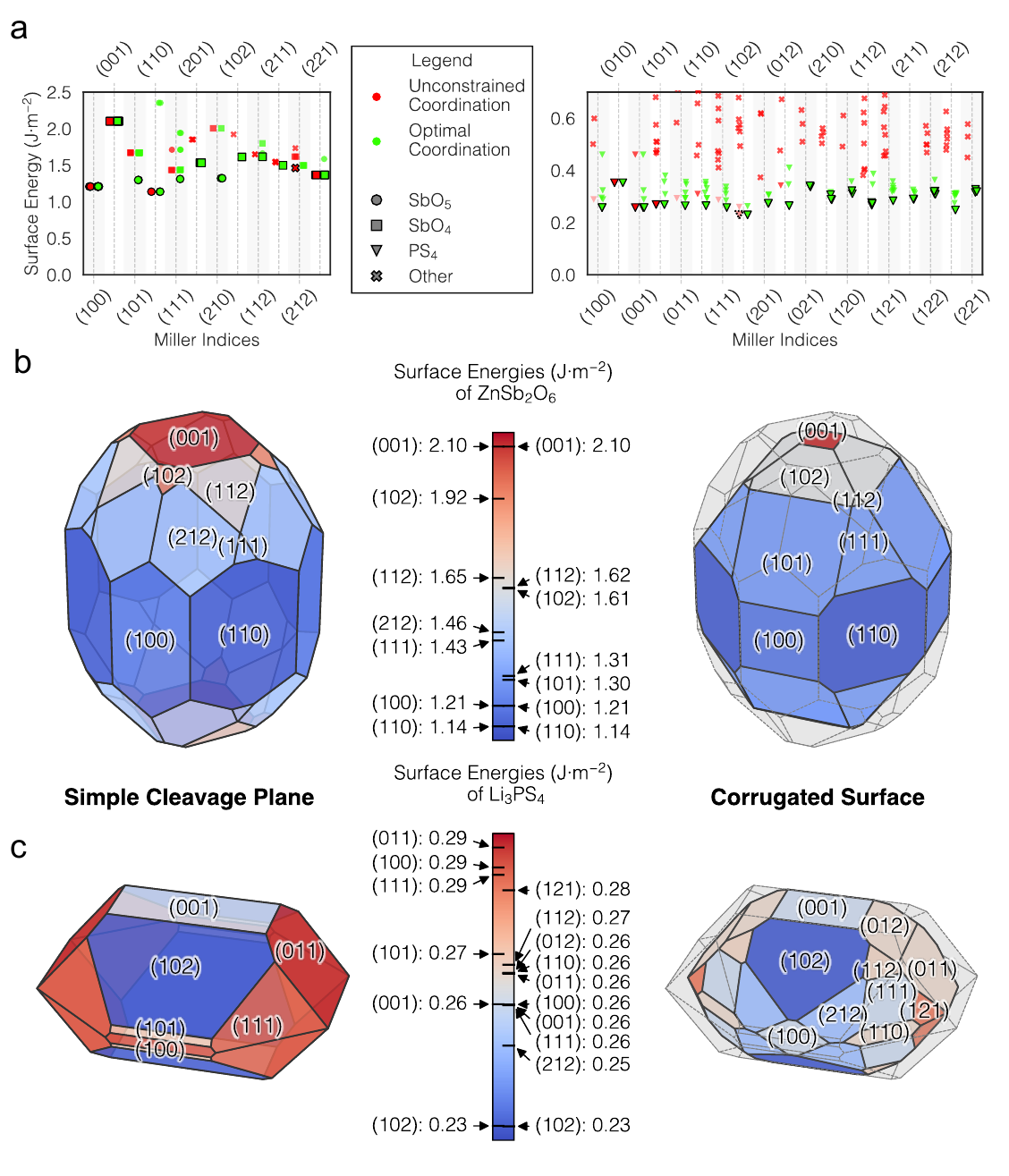}
\caption[Surface energies for constrained and unconstrained coordinated slabs]{
Comparison of surface energies and resulting Wulff shapes using unconstrained coordination versus constrained coordination.
\textbf{a}: Surface energies ($\gamma$) of slabs generated with unconstrained coordination (red) and constrained coordination (green) for \ce{ZnSb_2O_6} (left) and \ce{Li_3PS_4} (right). Faded red markers with dotted outlines indicate pseudo-planar configurations generated by the unconstrained coordination baseline (\autoref{sec:si:flat}). Marker shapes indicate the lowest primary coordination present at the surface.
\textbf{b, c}: Wulff shapes of \ce{ZnSb_2O_6} (\textbf{b}) and \ce{Li_3PS_4} (\textbf{c}) constructed from unconstrained coordination (colored polyhedra on the left, transparent outlines on the right) versus those from constrained coordination (solid colored polyhedra on the right). The equilibrium shapes predicted via constrained coordination are completely enclosed within the volumes generated by unconstrained coordination due to lower surface energies.
}
\label{fig:coordinatedvsundercoordinated}
\end{figure}

\autoref{fig:coordinatedvsundercoordinated}a presents the calculated surface energies ($\gamma$) across various Miller indices for \ce{ZnSb_2O_6} and \ce{Li_3PS_4}.
Multiple terminations are evaluated for each orientation, with the lowest energy configuration highlighted by opaque markers.
Except for one outlier---the \surfacemi{212} surface of \ce{ZnSb_2O_6}---the lowest surface energy is always achieved by surfaces satisfying the optimal coordination constraints.
Unconstrained coordination yields low energies only when it coincidentally preserves optimal surface coordination.

In \ce{ZnSb_2O_6}, without preserving the coordination environments of Sb, the \surfacemi{110} surface exhibits the lowest energy of 1.14~$\mathrm{J\cdot m^{-2}}$, constituting 27.0\% of the surface area of the Wulff shape (\autoref{fig:coordinatedvsundercoordinated}b, left).
The \surfacesmi{212} facets occupy most of the surface area (34.9\%), despite having a higher surface energy of 1.46~$\mathrm{J\cdot m^{-2}}$.
With constrained coordination, low-energy surfaces of the Wulff shape exhibit either \ce{SbO_5} or \ce{SbO_4} terminations.
The lowest-energy surface remains the \surfacemi{110} surface because the \ce{SbO_5} termination can be generated by unconstrained coordination.
The outlier \surfacesmi{212} facets (featuring \ce{SbO_3} termination) decompose into \surfacesmi{101} facets, which occupy 37.9\% of the total surface area (\autoref{fig:coordinatedvsundercoordinated}b, right) with an energy of 1.30~$\mathrm{J\cdot m^{-2}}$.

For \ce{Li_3PS_4}, preserving \ce{PS_4} units is necessary to achieve the minimum surface energy.
The global minimum surface energy across all generated slabs is 0.23~$\mathrm{J\cdot m^{-2}}$ on the \surfacemi{102} orientation, featuring fully intact \ce{PS_4} tetrahedra.
For slab models with initially broken \ce{PS_4} units, the lowest calculated energy is 0.37~$\mathrm{J\cdot m^{-2}}$ on the \surfacemi{201} surface.
In models generated via unconstrained coordination containing broken \ce{PS_4} units, structural relaxation only partially restores \ce{PS_3} units into \ce{PS_4}.
Therefore, spontaneous surface reconstruction is insufficient to achieve low-energy configurations.

In the Wulff shape derived from unconstrained coordination, the \surfacesmi{102} facets dominate, occupying 37.4\% of the total surface area (\autoref{fig:coordinatedvsundercoordinated}c, left).
Upon applying constrained coordination, the \surfacesmi{102} facets remain the lowest-energy facets, but their combined area contribution decreases to 23.0\%.
This area reduction results from the emergence of new low-energy orientations, producing a more multifaceted and isotropic Wulff shape.

Including the aforementioned pseudo-planar surfaces (\surfacemi{100}, \surfacemi{011}, \surfacemi{102}, and \surfacemi{111} orientations where the unconstrained coordination approach coincidentally preserves surface \ce{PS_4} units), the Wulff shape volume decreases by 19\% for \ce{ZnSb_2O_6} and 21\% for \ce{Li_3PS_4} when applying constrained coordination (\autoref{fig:coordinatedvsundercoordinated}b and c).
If these pseudo-planar surfaces are strictly excluded from the unconstrained baseline, the volume reduction of the Wulff shape for \ce{Li_3PS_4} becomes 46\% (\autoref{sec:si:flat}).

\section{Discussion}\label{sec:discussion}

In this work, we developed \pkgname{} to generate representative slab models for multinary compounds, which requires careful handling of stoichiometry, coordination environment, and slab symmetry.
To satisfy criteria of valid slabs, \pkgname{} evaluates all combinations of removing $n$ atoms from a pool of $m$ available surface atoms, scaling as $C_m^n = \frac{m!}{n!(m-n)!}$.
As the search space grows combinatorially with the number of available surface atoms ($m$) and required removals ($n$), \pkgname{} restricts the search region using a hyperparameter.
This parameter is defined as the ratio between the search depth and the longest atomic pair distance (\textit{e.g.}, \ce{Zn-Zn} bond of 4.7~\AA{}~in \ce{ZnSb_2O_6}).
By default, this ratio is set to 1.
If a valid slab is not found within this default surface region, increasing the search depth expands the configurational space.
However, there are limitations on scaling.
The size of the available atom pool $m$ is proportional to the area of the irreducible simulation cell of the slab model.
For studying low Miller index surfaces of materials with small primitive cells, $m$ is small enough to make the combinatorial space computationally trivial.
However, for surfaces with high Miller indices or bulk structures with large primitive cells, the exponential growth of the search space becomes the computational bottleneck.
\pkgname{} mitigates this through parallelization and rapid filtering, and we are working on further optimizations and enhanced algorithms to overcome the limitations of this approach.

When applying \pkgname{} to new materials, determining the optimal surface coordination environment \textit{a priori} is challenging, even with the guidance of bond strength analysis.
For instance, although the \ce{Sb-O} bonds in \ce{ZnSb_2O_6} are stronger than \ce{Zn-O} bonds, preserving intact \ce{SbO_6} octahedra at the surface is electrostatically prohibited.
Therefore, the practical application strategy for \pkgname{} involves an iterative exploration of the configurational space.
By generating slab models across a series of reasonable coordination criteria and evaluating their respective surface energies, the physical coordination limits for a given material can be systematically determined.

To reduce the initial configurational space, heuristic rules derived from structural prototypes can be extrapolated to similar multinary systems.
In solid-state electrolytes, ionic conduction relies on weaker bonding of mobile ions compared to the rigid structural framework.
Therefore, enforcing the geometric integrity of strongly covalent polyanions (\textit{e.g.}, \ce{PS_4^{3-}}, \ce{PO_4^{3-}}, or \ce{GeS_4^{4-}}) while relaxing the coordination criteria of the conducting alkali ions provides a computationally efficient and generalizable algorithm for generating low-energy surfaces in other fast-ion conductors.
Furthermore, aliovalent substitution with elements that form stronger covalent bonds changes the surface energies of different crystallographic orientations, which is a potential strategy to preferentially expose desired surfaces.

Explicit surface models of multinary trirutile oxides have so far been difficult to construct, and relative surface stabilities have therefore been assessed indirectly---for example, by transferring the sequence of surface energies of a simpler binary counterpart, such as rutile \ce{SnO_2}, when studying a trirutile \ce{ZnSb_2O_6} structure~\cite{jacksonComputationalPredictionExperimental2022}.
Our workflow removes this modeling limitation: by generating symmetric, stoichiometric, and optimally coordinated slabs for every symmetrically inequivalent orientation up to a Miller index of 2, the full surface-energy hierarchy of \ce{ZnSb_2O_6} becomes directly accessible.
Our result confirms that the \surfacemi{110} surface is indeed the lowest-energy surface, consistent with previous studies~\cite{jacksonComputationalPredictionExperimental2022,claesScreeningASb2O6Mg2026}.

After the baseline behavior of the stoichiometric surfaces is established, \pkgname{} can be extended to more complex defect chemistry.
While this study focuses on stoichiometric models, \pkgname{} natively supports the generation of off-stoichiometric surfaces by disabling the criterion of stoichiometry, which is a part of ongoing work.

In summary, by applying optimal coordination constraints, \pkgname{} makes accessible stable surface orientations that no planar cleavage can produce.
Replacing manual generation with an automated, surface-energy-guided methodology provides a reliable and transferable framework for high-throughput screening of surface properties across multinary materials.

\section{Methods}\label{sec11}

\subsection*{Structural models and Density functional theory calculations}
\noindent The initial bulk structure for $\beta$-\ce{Li_3PS_4} was based on the $Pnma$ structural model (ICSD-180319)~\cite{hommaCrystalStructurePhase2011}. Our previous study~\cite{xieEffectsGrainBoundaries2024} indicates that the lowest energy ordering is obtained by \ce{Li} atoms fully occupying the Li2 sites. This ordering was selected to generate slab models.
The ordered bulk structure for \ce{ZnSb_2O_6} was obtained from the Materials Project database (mp-3188)~\cite{jainCommentaryMaterialsProject2013}.

Density functional theory (DFT) calculations were performed using the Vienna \textit{ab initio} Simulation Package (VASP)~\cite{kresseInitioMolecularDynamics1993,kresseEfficientIterativeSchemes1996,kresseUltrasoftPseudopotentialsProjector1999}.
Core electrons were described by the projector-augmented wave method by Bl\"{o}chl~\cite{blochlProjectorAugmentedwaveMethod1994,kresseUltrasoftPseudopotentialsProjector1999}, and a cutoff of 520~eV is applied for the plane wave expansion~\cite{kresseEfficientIterativeSchemes1996}.

For the structural optimisation and the energy calculations of bulk and slab models, exchange-correlation interactions were described at the generalised gradient approximation (GGA) level~\cite{perdewGeneralizedGradientApproximation1996}.
The Perdew-Burke-Ernzerhof for solids (PBEsol) functional~\cite{perdewRestoringDensityGradientExpansion2008,perdewErratumRestoringDensityGradient2009} was selected for \ce{ZnSb_2O_6} and \ce{Li_3PS_4}.
To achieve higher accuracy in evaluating the electronic structure for crystal orbital Hamilton population analysis via LOBSTER, the hybrid PBE0~\cite{adamoReliableDensityFunctional1999} functional was used for the bulk structures.
Bader charge was calculated using the Bader Charge Analysis code~\cite{henkelmanFastRobustAlgorithm2006,tangGridbasedBaderAnalysis2009,sanvilleImprovedGridbasedAlgorithm2007,yuAccurateEfficientAlgorithm2011} on the charge density derived from the HSE06~\cite{heydHybridFunctionalsBased2003,heydErratumHybridFunctionals2006,paierScreenedHybridDensity2006,krukauInfluenceExchangeScreening2006} functional, considering the large cell size of the slab model.

During structural optimisation, the atomic positions were fully relaxed in all structures.
For the bulk structure, the lattice relaxation is allowed on all degrees of freedom (length and angle) without preserving the symmetry.
Dense $\Gamma$-centred $k$-point grids ensured convergence of DFT calculations within 5~$\mathrm{meV/atom}$.

\subsection*{Slab model generation and Wulff shape construction}
\noindent Slab models are generated using \pkgname{} package as described in \autoref{sec:workflow}.
Specifically, we evaluated all symmetrically inequivalent surface orientations with a maximum Miller index of 2, including 12 unique crystallographic orientations for \ce{ZnSb_2O_6} and 19 unique orientations for \ce{Li_3PS_4}.
For each given orientation, an exhaustive sampling of the cleavage plane position along the surface normal was performed to generate and evaluate multiple potential stoichiometric terminations.
The slab thickness is set to be at least 25~\AA{} to ensure that the central region of the slab is bulk-like.
However, during the atomic trimming process, the slab thickness can be slightly reduced.
A vacuum thickness of at least 15~\AA{} is applied to avoid interactions between periodic images.
The ratio between the search depth and the longest atomic pair distance of the material is set to 1.
For slab models, we fix all the lattice parameters to their respective bulk values.
The $k$-point grids of slab models are generated such that the density of $k$-points is comparable to that of bulk models, to ensure that the energies are comparable~\cite{sunEfficientCreationConvergence2013}.

The equilibrium crystal shape (Wulff shape) was constructed and visualized using the \texttt{WulffShape} module implemented in \texttt{pymatgen}.
The Wulff shapes derived from unconstrained versus constrained surface models were uniformly normalized using the same scale factor relating the surface energy to the distance from the center of mass for each material.

\backmatter

\bmhead{Acknowledgements}

\noindent W.X and P.C. acknowledge funding from the National Research Foundation under NRF Fellowship NRFF12-2020-0012.
W. X. and D.O.S. acknowledge the PRAETORIAN project, funded by UK Research and Innovation (UKRI) under the UK government's Horizon Europe funding guarantee (EP/Y019504/1).
We would like to acknowledge that the computational work involved in this research work is mainly supported by the National University of Singapore's IT Research Computing group under scheme NUSREC-HPC-00001 (\url{https://nusit.nus.edu.sg}).
We thank the support provided by Dr.\ Miguel Dias Costa and Dr.\ Wang Junhong.
A proportion of computational work was performed on resources of the National Supercomputing Centre, Singapore (\url{https://www.nscc.sg}).
Some computations in this work were performed using the University of Birmingham's BlueBEAR HPC service, which provides a High Performance Computing service to the University's research community (\url{http://www.birmingham.ac.uk/bear}).
Some calculations were performed using the Sulis Tier 2 HPC platform hosted by the Scientific Computing Research Technology Platform at the University of Warwick (funded by EPSRC Grant EP/T022108/1 and the HPC Midlands+ consortium).
Through our membership of the UK's HEC Materials Chemistry Consortium, which is funded by the UK Engineering and Physical Sciences Research Council (EPSRC; EP/L000202, EP/R029431, and EP/T022213), this work also used ARCHER2 UK National Supercomputing Services.
The authors are grateful to the UK Materials and Molecular Modelling Hub for computational resources, which is partially funded by EPSRC (EP/T022213/1, EP/W032260/1, and EP/P020194/1).
This work used computational resources of the supercomputer Fugaku provided by the RIKEN Centre for Computational Science under the ``Fugaku Projects via National Supercomputing Centre Singapore'', and through the HPCI System Research Project (Project ID: hp230188).

\bmhead{Competing interests}

\noindent The authors declare no competing interests.

\bmhead{Data availability}

\noindent The calculation data, including bulk structures and slab structures before and after relaxation, are available at \url{https://github.com/caneparesearch/project_SALAMI}. Additional data is available upon request.

\bmhead{Code availability}

\noindent The \pkgname{} code is open source at \url{https://github.com/caneparesearch/salami}. The API documentation is available at \url{https://salami.readthedocs.io/}.

\bmhead{Author contributions}

\noindent W.X. designed the codebase, generated the surface models, performed all density functional theory calculations, and wrote the original manuscript draft. Z.L. proposed the necessary geometric conditions for generating symmetric planes. P.C. acquired funding and conceptualized the initial problem. All authors discussed the results, contributed to the interpretation, and reviewed the final manuscript.

\begin{appendices}

\section{Coordination constraints}
\noindent The initial coordination constraints applied during the intermediate stages of the \pkgname{} workflow for \ce{ZnSb_2O_6} (\autoref{code:fig:slabworkflow}) are summarized in \autoref{tab:coordinationworkflow}.
\begin{table*}[!htbp]
 \centering
 \small
 \caption{Initial coordination constraints of \ce{ZnSb_2O_6} for the demonstration of the \pkgname{} workflow.}
 \label{tab:coordinationworkflow}
 \begin{tabular*}{\textwidth}{@{\extracolsep{\fill}}lcc@{}}
 \toprule
 \textbf{Bond Type} & Min CN & Max CN \\
 \midrule
 \ce{Zn-O} & 1 & 6 \\
 \ce{Sb-O} & 5 & 6 \\
 \ce{O-(Sb/Zn)} & 1 & 3 \\
 \bottomrule
 \end{tabular*}
\end{table*}

The optimal coordination criteria for \ce{ZnSb_2O_6} and \ce{Li_3PS_4} are shown in \autoref{tab:coordination}.
The max coordination number (CN) is the bulk CN.
The coordination criteria from the min CN ensure that there is no dangling bond, \textit{i.e.}, isolated atom in vacuum.
\begin{table*}[!htbp]
 \centering
 \small
 \caption{Coordination numbers for scenarios of optimal constraints and unconstrained coordination.}
 \label{tab:coordination}
 \begin{tabular*}{\textwidth}{@{\extracolsep{\fill}}lcccc@{}}
 \toprule
 \textbf{Bond Type} & \multicolumn{2}{c}{\textbf{Constrained Coordination}} & \multicolumn{2}{c}{\textbf{Unconstrained Coordination}} \\
 \cmidrule(lr){2-3} \cmidrule(lr){4-5}
 & Min CN & Max CN & Min CN & Max CN \\
 \midrule
 \ce{Zn-O} & 1 & 6 & 1 & 6 \\
 \ce{Sb-O} & 4 or 5 & 6 & 1 & 6 \\
 \ce{O-(Sb/Zn)} & 1 & 3 & 1 & 3 \\
 \midrule
 \ce{Li-S} & 1 & 4 & 1 & 4 \\
 \ce{P-S} & 4 & 4 & 1 & 4 \\
 \ce{S-(P/Li)} & 1 & 5 & 1 & 5 \\
 \bottomrule
 \end{tabular*}
\end{table*}

\section{Software Implementation and Energy Filtering}\label{sec:package}

\noindent Designed with cross-platform compatibility across Windows, macOS, and Linux environments, \pkgname{} is an open-source \texttt{Python} package built compatible with the \texttt{pymatgen} library~\cite{ongPythonMaterialsGenomics2013,tranSurfaceEnergiesElemental2016,sunEfficientCreationConvergence2013,shinoharaAlgorithmsMagneticSymmetry2023,togoSpglibSoftwareLibrary2024}.
The API documentation of \pkgname{} is available at \url{https://salami.readthedocs.io/en/latest/}.
The package is designed to automate the generation, evaluation, and energetic pre-screening of multinary surface models.

The core surface generation is executed by the \texttt{SlabTrimmer} class within the \texttt{\pkgname{}.generator} module.
Starting from an initially asymmetric slab generated via a cleavage plane, the algorithm systematically trims atoms to produce a valid structural model.
The criteria for a valid slab are defined in the \texttt{\pkgname{}.evaluator} module.
By default, the routine requires the slab to be symmetric (to yield unambiguous surface energies, $\gamma$), nonpolar (to eliminate surface dipoles), and optimally coordinated (to preserve high-energy bonds).
Stoichiometry and charge neutrality are also enforced by default, though users can relax these constraints to model off-stoichiometric reconstructions under varying chemical potentials.

Applying these constraints---particularly the restoration of stoichiometry through combinatorial atom removal---generates a large pool of candidate configurations.
Because the configurational space scales factorially with the number of active surface atoms, evaluating all candidates directly via DFT is computationally impractical.
To mitigate this, the \texttt{\pkgname{}.filter} module is implemented to pre-screen configurations using computationally inexpensive energy evaluations.
As a general-purpose tool, \pkgname{} supports electrostatic evaluations via Ewald summation~\cite{toukmajiEwaldSummationTechniques1996}, empirical potentials through a \texttt{LAMMPS}-\texttt{Python} integration~\cite{thompsonLAMMPSFlexibleSimulation2022}, and provides flexibility for user-defined machine-learning interatomic potentials.
The simple Ewald summation was used in this study.

Based on these approximate energies, the configurational space is narrowed down using user-defined selection strategies.
The standard approach is to rank the slab models by their energies and keep the lowest-energy configurations for subsequent DFT relaxation.
Alternatively, configurations can be sampled stochastically using a Boltzmann distribution, which is useful for studying finite-temperature effects.
Finally, to accelerate the exhaustive combinatorial search, the evaluation process---specifically, verifying whether a given trimming combination produces a valid slab that satisfies all predefined criteria---is parallelized and distributed across multiple CPU cores.

\section{Pauling's Rules at the Surface}\label{sec:si:pauling}
\noindent To understand the intrinsic structural limitations at the surface of \ce{ZnSb_2O_6}, we apply Pauling's electrostatic valence principle.
The electrostatic bond strength ($s_{ij}$) is defined by the ratio of the cation's formal oxidation state ($V_i$) to its coordination number ($N_i$):
\begin{equation}
s_{ij} = \frac{V_i}{N_i}.
\end{equation}

In the bulk phases of \ce{ZnSb_2O_6}, we consider \ce{Sb^{5+}} and \ce{Zn^{2+}} in octahedral sites ($N=6$):
\begin{itemize}
 \item Bond valence for \ce{Sb-O}: $s_{\text{Sb}} = \frac{5}{6}\approx 0.83$ v.u.
 \item Bond valence for \ce{Zn-O}: $s_{\text{Zn}} = \frac{2}{6}\approx 0.33$ v.u.
\end{itemize}

Each \ce{O} atom is coordinated by two \ce{Sb} and one \ce{Zn} atom, satisfying the valence sum ($V_{\text{O}}=-2$):
\begin{equation}
\sum s_{ij} = 2 \times \frac{5}{6} + \frac{1}{3} = 2 \text{ v.u.}
\end{equation}

Suppose an ideal, stoichiometric surface is created where the \ce{SbO_6} octahedra remain intact.
At least one \ce{O} atom per \ce{Sb} unit must point toward the vacuum (the ``outer'' oxygen), losing its connectivity to the rest of the lattice.
In every formula unit of \ce{ZnSb_2O_6}, there are two surface \ce{SbO_6} units, and therefore, two such undercoordinated \ce{O} atoms.
The valence deficiency ($\Delta$) for a single such \ce{O} atom, assuming it remains bonded only to its \ce{Sb} center, is:
\begin{equation}
\Delta = |-V_{\text{O}}| - s_{\text{Sb}} = 2 - 0.83 = 1.17\text{ v.u.}
\end{equation}
For the two surface \ce{SbO_6} units, the cumulative deficiency is $2 \times 1.17 = 2.33\text{ v.u.}$

To restore electrostatic neutrality and satisfy the local charge balance for these undercoordinated \ce{O} atoms, the single \ce{Zn^{2+}} cation available at the surface (per formula unit) must compensate for this deficiency.
However, a single \ce{Zn^{2+}} cation provides a maximum of only $2\text{ v.u.}$ to distribute.
Since $2.33\text{ v.u.} > 2\text{ v.u.}$, it is impossible for the stoichiometrically available \ce{Zn} to satisfy the valence requirements of the ``outer'' \ce{O} atoms while keeping the \ce{Sb} hexacoordinated.
Therefore, the system must reduce the coordination number of \ce{Sb} at the surface.

\section{Bader Charge Analysis of \ce{ZnSb_2O_6} \surfacemi{212} surface}\label{sec:si:bader}

\setcounter{table}{0}

The calculated Bader charges and inferred oxidation states for atoms in the bulk structure and at the \surfacemi{212} surface of \ce{ZnSb_2O_6} are shown in \autoref{tab:bader_charges}.

\begin{table*}[!h]
 \centering
 \small
 \caption{Calculated Bader charges and inferred oxidation states for bulk and \surfacemi{212} surface atoms in \ce{ZnSb_2O_6}.}
 \label{tab:bader_charges}
 \begin{tabular*}{\textwidth}{@{\extracolsep{\fill}}llcc@{}}
 \toprule
 \textbf{Structure} & \textbf{Atom} & \textbf{Bader Charge ($|e|$)} & \textbf{Inferred State} \\
 \midrule
 \textbf{Bulk} & \ce{Zn} & $+1.3 \sim +1.4$ & $\sim \ce{Zn^{2+}}$ \\
 & \ce{Sb} & $+2.7 \sim +2.9$ & $\sim \ce{Sb^{5+}}$ \\
 & \ce{O} & $-1.2 \sim -1.1$ & $\sim \ce{O^{2-}}$ \\
 \midrule
 \textbf{\surfacemi{212} surface} & \ce{Sb} (Tri-coordinated) & $+1.8$ & $\sim \ce{Sb^{3+}}$ \\
 & \ce{O} (Oxygen dimer) & $-0.5$ & $\sim \ce{O^-}$ \\
 \bottomrule
 \end{tabular*}
\end{table*}

\section{Pseudo-planar \ce{Li_3PS_4} surfaces}\label{sec:si:flat}

\begin{figure}[h!]
 \centering
 \includegraphics[width=0.5\textwidth]{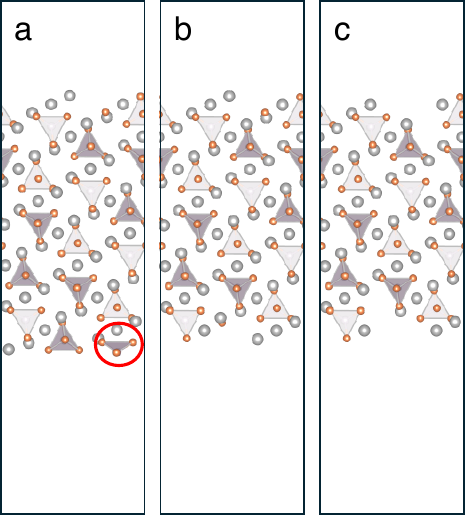}
 \caption{Generation of the pseudo-planar \surfacemi{011} surface.
 \textbf{a}: The initial asymmetric slab generated via a cleavage plane, containing a broken \ce{PS_3} unit.
 \textbf{b}: The intermediate slab after symmetrization, where the broken unit on the bottom surface is removed.
 \textbf{c}: The final symmetric, stoichiometric slab after surface atom trimming.}
 \label{fig:lps011}
\end{figure}

As noted in \autoref{sec:undercoordination}, while it is impossible to generate intact \ce{PS_4} units via a raw planar cleavage along the \surfacemi{100}, \surfacemi{011}, \surfacemi{102}, and \surfacemi{111} orientations (\textit{e.g.}, \surfacemi{011} as shown in \autoref{fig:cuttingpath}), the unconstrained coordination baseline appears to successfully generate intact \ce{PS_4} units on these orientations.
This occurs because both the constrained and unconstrained slab models undergo the identical symmetrization pipeline in \pkgname{} (Step 2 of \autoref{code:fig:slabworkflow}).
For example, generating the \surfacemi{011} surface initially yields an asymmetric and stoichiometric slab via \texttt{pymatgen}, which inevitably contains a broken \ce{PS_3} unit, as shown in \autoref{fig:lps011}a.

When \pkgname{} symmetrizes this slab using the upper surface as the reference, it trims atoms from the bottom surface. This step coincidentally removes the broken \ce{PS_3} unit (\autoref{fig:lps011}b).
The subsequent stoichiometry recovery removes additional atoms symmetrically from both sides, resulting in a final slab where all remaining \ce{P} atoms are fully coordinated as \ce{PS_4} (\autoref{fig:lps011}c).

\begin{figure}[h!]
 \centering
 \includegraphics[width=1\textwidth]{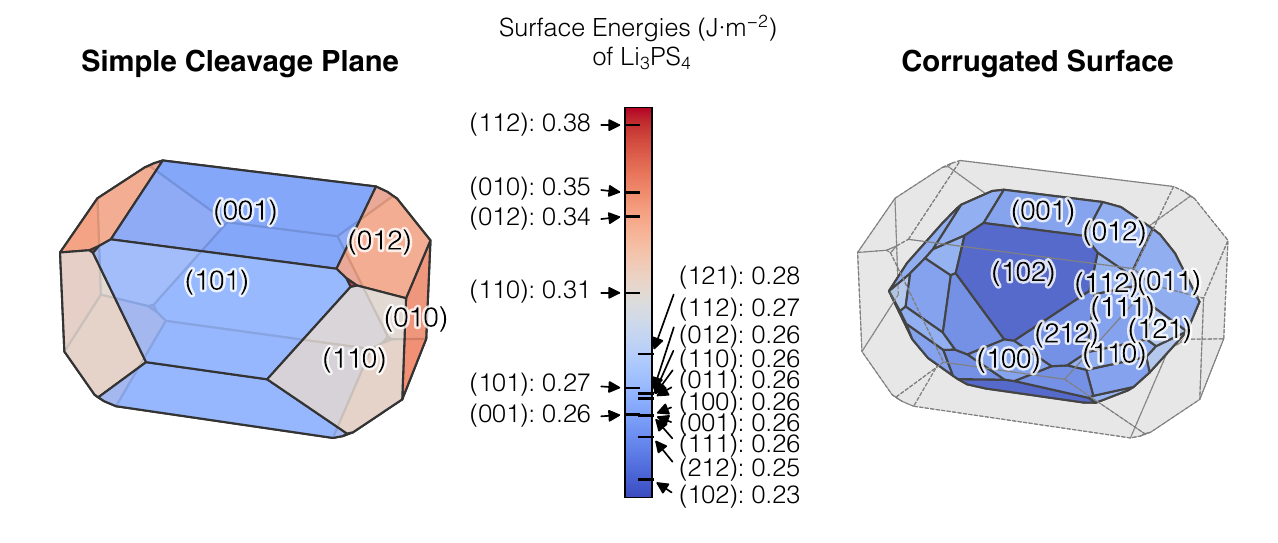}
 \caption{Wulff shape of \ce{Li_3PS_4} generated by excluding the \surfacemi{100}, \surfacemi{011}, \surfacemi{102}, and \surfacemi{111} surfaces, where a planar cleavage inevitably breaks \ce{PS_4} units.
 }
 \label{fig:wronglpswulff}
\end{figure}

If these \surfacemi{100}, \surfacemi{011}, \surfacemi{102}, and \surfacemi{111} surfaces with intact \ce{PS_4} units are excluded, the resulting Wulff shape is shown in \autoref{fig:wronglpswulff}.
The volume of the correct Wulff shape (\autoref{fig:coordinatedvsundercoordinated}c, right), where \ce{PS_4} units are strictly preserved across all surfaces using constrained coordination, is 46\% smaller than the volume of this Wulff shape in \autoref{fig:wronglpswulff}.

\end{appendices}

\bibliography{cited}

\end{document}